# Use Cases for Prospective Sensemaking of Human-AI-Collaboration

Ishara Sudeeptha
Victoria University of Wellington
ishara.kandanearachchigedon@vuw.ac.nz

Wieland Müller
University of Rostock
wieland.mueller@uni-rostock.de

Michael Leyer
University of Marburg
Queensland University of Technology
michael.leyer@wiwi.uni-marburg.de

Alexander Richter
Victoria University of Wellington
alex.richter@vuw.ac.nz

Ferry Nolte
Leibniz University Hannover
nolte@iwi.uni-hannover.de

**Abstract**

*This study explores the potential of Human-AI Collaboration (HAIC) use cases as a tool for prospective sensemaking. Based on 14 interviews with executives of an automotive company, we identify and categorize HAIC use cases that can help organizations anticipate and strategically respond to the impact of HAIC. Feedback from the case company shows that our systematic mapping of HAIC use cases along the value chain and group tasks enables a structured understanding of the potential role of AI and underscores the importance of strategic foresight when integrating AI into organizational processes.*

## 1. Introduction

The speedy adoption of ChatGPT across many organizational practices illustrates significant benefits of adopting methods that engage with future possibilities. Organizations can gain substantial advantages from these forward-looking approaches by engaging in a foresight process, where prospective sensemaking plays a crucial role in shaping this future-oriented perspective (Tapinos & Pyper, 2018). IS research can contribute to this by applying a forward-looking perspective when designing and developing new IS artefacts. One of the methods in question is prospective sensemaking, which involves anticipating potential developments and identifying influential factors in the face of unpredictability and ambiguity (Gattringer et al., 2021; Wright, 2005). It helps exploring rapidly emerging technologies and their broader potential impact.

This is particularly true for HAIC. While many organizations are currently exploring its potential through various use cases to support their work, it remains unclear how HAIC can be optimally implemented and utilized across different organizational functions (Wamba-Taguimdje et al., 2020). Effective utilization of HAIC requires a comprehensive approach that considers the entire value-creating activities in an organization and understanding the various group tasks required for successful HAIC. The identification of existing and potential HAIC use cases highlights the areas where AI integration is perceived as valuable. These use cases therefore not only highlight the practical applications of AI but also demonstrate its strategic value in improving organizational activities and decision-making processes. This, in turn, foster a better understanding and strengthens the current and prospective role of HAIC. Hence, we ask: *How can HAIC use cases be identified and classified to support prospective sensemaking?*

We employed a case study approach, mainly drawing from 14 semi-structured interviews with executives, to identify HAIC use cases that can support prospective sensemaking. Furthermore, we used Porter's (1985) value chain model and McGrath's (1984) group task circumplex model to create a novel matrix that further classifies the identified use cases.

This study contributes by highlighting the potential of AI use cases as a valuable tool for prospective sensemaking in the context of HAIC. The identification of AI use cases enables organizations to anticipate the impact of emerging technologies, facilitating informed decision-making and strategic planning for AI adoption and implementation. The research further contributes by developing and applying a novel holistic framework that classifies HAIC use cases within organizations by integrating Porter's value chain model and McGrath's group task circumplex model.

## 2. Prospective Sensemaking for HAIC

### 2.1. Conceptualization

In this study, we suggest that organizations engage in prospective sensemaking activities focused on future-oriented technological developments, guided by present circumstances characterized by rapid technological advancements. For example, sectors like the automotive industry increasingly integrate AI into their products and enhance efficiency in production processes. Along with technological advancements, the competitive landscape plays a critical role, where organizations not only experiment with AI but also successfully implement it. Hence, organizations need to accelerate their adoption efforts to remain competitive. By understanding these present circumstances, organizations can develop a more nuanced understanding of potential technologies and their impact on users. By doing so, organizations can anticipate unfamiliar situations and develop suitable interpretations proactively (Friesl et al., 2019). Prospective sensemaking opens a new avenue for navigating the untapped terrains of emerging and evolving technologies. This approach allows researchers to anticipate impacts and make preliminary decisions that help in better navigating technological changes to achieve success (Schneckenberg et al., 2017).

It is also crucial to understand how prospective sensemaking differs from traditional sensemaking and why it is beneficial as an approach for future-oriented research in IS. Traditional sensemaking is more retrospective, used to understand current or past crises or failures (Klein et al., 2007; Weick, 1993). In contrast, prospective sensemaking is future-oriented and forward-looking (Tapinos & Pyper, 2018). With the rapid development and widespread adoption of AI, there is an increasing focus on identifying the future impact of these technologies (Makridakis, 2017). In this context, prospective sensemaking becomes critically important to anticipate the consequences and impacts of these technologies (Gattringer et al., 2021).

Weick (1995) identifies the noticing of relevant items and concepts, as well as the grouping and acting on use cases, as central properties of sensemaking in organizations. Our study extends this process toward prospective sensemaking, which is forward-looking, while recognizing the importance of the aforementioned activities in sensemaking. A broad analysis of potential developments and use cases contributes to a richer selection of perspectives in the sensemaking process (Gattringer et al., 2021). Therefore, it is crucial to identify and group HAIC use cases in organizations to understand the sensemaking process, align organizational strategy with the sensemaking of individuals, and optimize HAIC collaboration.

In this study, we argue that the identification and grouping of HAIC use cases are vital in anticipating the consequences of emerging technologies within the context of HAIC. Structuring these use cases allows organizations to delve deeper into exploring potential, ultimately enhancing the alignment of their strategic decisions with anticipated technological advancements related to HAIC. This prospective sensemaking approach is particularly important in the field of information systems, where the integration of AI-related technologies is on the rise (Gattringer et al., 2021). This process not only enables organizations to develop strategies for future HAIC integration but also provides insights into the implications for information systems research, particularly in areas such as human-AI group tasks, changes in organizational value-adding activities, and organizational structures. Thus, prospective sensemaking serves as a critical process in preparing organizations for emerging technological advancements by making their strategies forward-looking, thereby ensuring successful HAIC efforts.

This study highlights the importance of anticipating the potential impact of Human-AI Collaboration (HAIC) across various organizational activities and group tasks through a prospective sensemaking approach. We engaged a diverse group of executives in the case company and conducted semi-structured interviews, through which we identified and categorized HAIC use cases across the value chain and group task circumplex activities. This systematic mapping of HAIC use cases provides a holistic and clear overview of the potential role AI can play in the context of HAIC from the executives' perspective. This prospective sensemaking approach equips organizations with a structured understanding of HAIC use cases. It also helps identify potential risks associated with some of these use cases. By doing so, organizations can prioritize HAIC use cases within the value chain and group tasks, focusing on areas where significant benefits can be realized.

### 2.2. Related Work

Implementing AI solutions also involves overcoming unexpected challenges and establishing reliable meanings about technology and data. Dolata and Crowston (2024) argue that the sensemaking process is integral to addressing these challenges and ensuring the successful deployment of AI technologies. This involves a continuous cycle of interpretation and adaptation, which is essential for maintaining the relevance and efficacy of AI solutions.

Collaboration in sensemaking is another critical aspect of the effective integration of AI. Vallis et al. (2023) highlight the necessity for individuals to engage in collaborative sensemaking, combining human understanding with AI-generated insights. This collaboration ensures that AI implementations are not only innovative but also practically effective, as diverse perspectives contribute to a more comprehensive understanding of AI capabilities and limitations.

Prospective sensemaking of AI is essential for shaping employees' adaptation within the workplace by involving them in interpreting the potential impact of AI technologies on their work and work environment (Weber et al., 2024). This process is crucial for identifying the uncertainties they may face in Human-AI Collaboration (HAIC) scenarios and for navigating these uncertainties through exploratory, proactive, or cautious approaches. By doing so, employees can better align their actions to collaborate effectively with AI and adapt to the evolving role of AI in the workplace.

The role of AI in design and innovation further illustrates the shift towards sensemaking. Verganti et al. (2020) argue that as AI increasingly takes over creative problem-solving tasks, human designers are tasked with understanding which problems should be addressed. This shift transforms design into an activity centered on sensemaking, focusing on interpreting and defining the problems that AI should solve.

Furthermore, Leyer and Schneider (2021) explore how AI has emerged as a promising technology for managerial decision-making, posing both opportunities and threats. Their research reveals that organizations can use AI-enabled decision-making solutions in various ways, each with implications for managerial job design. They discuss how AI can serve as tools or novelties, for decision augmentation or automation, and the outcomes related to these combinations. This highlights the importance of sensemaking in understanding and managing the interaction between human managers and AI technologies, ensuring that capabilities, responsibilities, and acceptance are effectively balanced.

## 3. Method

### 3.1. Case Study

Our empirical study is based on a case study of MULTI, an automotive original equipment manufacturer (OEM). MULTI is active in two main business areas: Automotive and Chemical Products. The company generates sales of over 40 billion euros and employs around 200,000 people at more than 50 locations worldwide. Its customer base includes global OEMs and regional retailers. MULTI operates within a hierarchical structure that ensures clear responsibilities but can hinder cross-functional collaboration. With a global presence, MULTI leverages diverse talent and markets, although this also complicates coordination and data management—areas where AI plays a crucial role.

The automotive sector is experiencing significant technological advances, particularly in autonomous driving, which integrates AI into products. Conversely, the chemical sector is using AI to improve the monitoring and efficiency of manufacturing and quality processes. MULTI has observed that its competitors are not only experimenting with AI but also using it successfully, forcing MULTI to accelerate its AI adoption to remain competitive. This strategic direction is supported by the internal corporate culture, which is characterized by values such as "freedom to act" and "passion to win" and empowers employees to innovate with AI technologies.

MULTI's approach to developing AI technologies is a balance between internal capabilities and external collaborations, particularly for components such as camera systems. The decision to develop AI technologies internally is driven by the need to protect intellectual property and control sensitive information in order to maintain a competitive advantage. The company is currently conducting several initiatives to evaluate the successful integration of AI. One of these is a pilot study with 800 users to integrate large language models, similar to ChatGPT, into daily operations to improve business processes. The company is also testing large-scale language models for internal purposes, such as reviewing supplier contracts and user agreements, and an AI-powered camera system for visual inspection of routine processes.

Office workers are very interested in integrating AI tools such as ChatGPT into their everyday work to automate tasks. However, MULTI faces major challenges in terms of data sensitivity and compliance. Strict rules apply to the types of data that can be entered into applications, especially for non-encrypted databases. For example, financial data is considered highly sensitive and must not be used in external, cloud-based AI applications. Despite the willingness of employees to use AI applications, there is a clear need for rules and guidelines to ensure compliance with data protection regulations. To overcome these challenges, MULTI is actively seeking employee feedback as part of a pilot study to conduct a needs analysis for AI applications.

This case study is part of a larger research initiative examining the opportunities and implications of AI implementation within MULTI. The broad-based pilot study is intended to take a comprehensive look at AI and examine various areas of application and implementation options of AI. The company wants to determine whether it should introduce a third-party AI tool company-wide or develop customized, internally developed AI solutions. The company also emphasizes the need for its own AI strategy. To this end, an organizational task force of experts, the works council, and senior management has been created to structure AI efforts and provide a framework for company members. This strategy will clearly define the permitted activities through AI and outline the allocated budgets for these initiatives.

One of the authors of this paper is employed by the case company and is actively involved in the pilot project to introduce HAIC. To avoid biases, the author only assisted with the data collection and was involved in the discussion of the study results, but not in the data analysis. The data analysis was conducted by two independent researchers to ensure objectivity.

### 3.2. Semi-structured Interviews

Interviews were chosen for this study to understand the HAIC use cases within the empirical case study. This method is particularly effective for researching complex topics. Semi-structured interviews offer a flexible yet structured approach, where the interviewer follows a guideline but may deviate from it depending on the course of the interview. This approach allows for in-depth insights while keeping the focus on the research objectives (Wilson, 2014).

Semi-structured interviews were conducted with executives from the case study company. The interview guide, available in full in the data repository, contains questions about the participants' roles, their understanding of AI and the expected impact on their work. The guide explores how AI could impact future capabilities, strategic planning, and organizational cultural adjustments. Each interview concludes with discussions about potential future use cases of HAIC within the organization and personal perspectives on the role of AI in their work environment.

A total of 14 interviews were conducted online via video call to capture a wide range of experiences and insights from different leadership positions and departments. The interview data is also used for another publication for which, however, different parts of the interviews are used. Detailed descriptions of the interviewees' roles and backgrounds are available in the data repository. The individuals identified were on a high management level; some of them focused on a specific functional area, but some of them were on a high management level so that all functional areas, according to Porter, were covered. The number of interviews was primarily based on the saturation point at which no new insights emerged - an important aspect emphasized in existing guidelines (Anderson, 2010).

Data analysis for the empirical case was conducted using NVivo 14, a leading qualitative analysis software that supports robust data management and helps to identify emerging patterns in rich narrative data. Two researchers conducted Initial coding independently to improve analytic rigor, integrate multiple perspectives, and reduce individual bias. Each coder developed independent coding schemes based on their interpretations of the data, which were later synchronized to agree on a unified coding structure. This approach ensures comprehensive data analysis while mitigating the effects of individual coder bias.

### 3.3. Categorization of HAIC Use Cases

After conducting and analyzing the semi-structured interviews, the identified HAIC use cases were systematically categorized using two established frameworks: Porter's Value Chain (Porter, 1985) and McGrath's Group Task Circumplex (McGrath, 1984). This approach provides a comprehensive understanding of how AI is applied within the organization, impacting both operational activities (value chain) and collaborative tasks (group task circumplex). The value chain categorization identifies how each HAIC use case supports different functional areas, while the group task circumplex categorization reveals the specific types of collaborative tasks involved in each use case.

To ensure the robustness and accuracy of this categorization, the use cases were first classified independently by three of the authors based on their interpretation of the data and the definitions in the frameworks. The results were then cross-checked in a joint discussion to reconcile any discrepancies and reach consensus on the final categorization. This method of independent categorization followed by subsequent voting helps to reduce individual bias and integrate multiple perspectives.

First, we categorized HAIC use cases by value chain activities to identify their support for organizational functions and then by group task circumplex activities to explore human-AI collaboration in group tasks. These frameworks were then integrated into a matrix, providing a holistic view of HAIC use cases that support both functional areas and group tasks across value-adding activities. This approach supports prospective sensemaking by illustrating AI's current value, identifying potential areas for future development, and strategically implementing AI within the organization. The methodical categorization clarifies HAIC use cases across value-adding activities and their support for group tasks, offering potential actionable insights to enhance future operational efficiency and foster HAIC.

Michael Porter's value chain model provides a framework for analyzing a company's various value-adding activities, classified as primary and supporting activities, and how they collectively contribute to the company's competitive advantage (Porter, 1985). Primary activities are value-creating and include inbound logistics, operations, outbound logistics, marketing and sales, and service. Supporting activities, such as firm infrastructure, human resource management, technological development, and procurement, provide essential support to the primary activities. In the modern world, the value chain model remains highly relevant, as breaking down these activities helps identify the novelty of innovations and enables organizations to integrate advanced technologies, such as data analytics, to enhance overall efficiency. (Nagy et al., 2018). In our study, the classification of use cases across the value chain supports prospective sensemaking by identifying the potential of HAIC use cases to support a wide array of functions across the organization.

McGrath's group task circumplex model can be used to analyze group dynamics and task typologies, offering a framework to understand diverse tasks and the interaction between task requirements and processes (McGrath, 1984). The vertical axis contrasts cooperative tasks requiring consensus with those involving conflict or negotiation. Cooperation, crucial for tasks with high interdependence, fosters mutual understanding and support. The horizontal axis differentiates conceptual tasks, focused on ideas, from behavioral tasks, which are action oriented. The model aids in optimizing group performance, designing collaboration tasks, and enhancing communication strategies, especially for tasks requiring consensus and coordination (Saavedra et al., 1993). In our study, we primarily focused on four quadrants in the circumplex: generate (tasks focused on generating new ideas), choose (tasks focused on selecting among alternatives), negotiate (tasks requiring conflict resolution and compromise), and execute (tasks related to implementation and actions). These quadrants are highly relevant when AI and humans collaborate to achieve group tasks. This classification supports prospective sensemaking by helping to understand the different use cases of HAIC that support various group tasks.

## 4. Results

We identified 63 potential use cases of HAIC through our case study analysis, where the organization can use them to support prospective sensemaking. These use cases were then categorized in a novel matrix, as demonstrated in Table 1. This matrix was developed by combining value chain activities (represented as rows) and group task circumplex activities (represented as columns).

Out of the 63 identified HAIC use cases, we identified 7 that were flagged by participants in the interviews as potentially riskier compared to others. Recognizing these risk factors is crucial, as it ensures that the classification not only highlights use cases with high potential benefits but also considers those with potentially higher risks (while other HAIC use cases may carry lower risks). By providing insights into these risks, the study offers a more balanced perspective for prospective sensemaking. This helps organizations make informed decisions by prioritizing HAIC use cases that deliver substantial potential value while mitigating potential risks and adverse impacts.

In the following section, we discuss a selection of use cases that serve as exemplary examples due to their high potential and significant risk. While these cases highlight critical areas of concern, we focus on only a few of them to illustrate the importance of cautious adoption. These selected use cases demonstrate the need for organizations to approach their potential implementation strategically, balancing the substantial benefits against the inherent risks.

Interview participants expressed mixed feelings about using AI for email management. While they identified potential benefits of AI managing emails, they also noted it as potentially risky due to doubts about AI's ability to align with their personalized work styles and its inefficiency in this context. Additionally, participants have personal communication preferences, prioritizing decision-making over mere efficiency in clearing out the inbox. Hence, they perceive that there would be little added value from an AI assistant managing the inbox.

Using AI to manage administrative tasks raises several privacy issues related to handling personal information. AI is quite helpful in managing requests related to expenditure and account details. However, when it comes to people-related businesses involving personal or proprietary information, AI raises significant privacy concerns. The involvement of AI in these areas may impact privacy laws, and therefore, AI should not be used in handling such sensitive information.

Conflict resolution is another area that presents potential risks. AI is perceived as inadequate in handling these complex tasks, as these heavily rely on human judgments that compare many competing demands. Additionally, AI cannot understand individual communication styles and personal nuances required in these contexts.

Interview participants mentioned that AI should be approached with caution if it is used in performance rating and evaluation. AI has several limitations in understanding human emotions involved in people management. While AI might automate up to 80% of performance evaluations, the remaining 20% should be considered irreplaceable, as AI cannot replicate the personal level and contextual understanding required to give feedback. Although AI can collect information from emails and team meetings to generate feedback, it lacks the personal touch and insight that a human supervisor provides.

**Table 1. Identified HAIC use cases and their assignment to group tasks and the value chain**

|  | Generate | Choose | Negotiate | Execute |
|---|---|---|---|---|
| **Firm infrastructure** | • Calendar management<br>• Managing resilience and economic shocks<br>• Meeting arrangement assistance | • Big data analytics<br>• Consolidation and summarization<br>• Create graphs and analyze data<br>• Data support for decision-making<br>• Knowledge management and Knowledge transfer<br>• Leadership and management<br>• Querying Knowledge Databases<br>• Decision-making and innovation |  | • Creating Presentations<br>• Documentation<br>• Email management<br>• Meeting summaries and reviews<br>• Managing administrative tasks |
| **Human Resource Management** | • Flexible scheduling for employees | • Categorization of HR work request tickets<br>• Insight and analytics from HR data<br>• Job evaluation<br>• Policy and process assistance<br>• Salary planning<br>• Performance rating and evaluation<br>• Providing supervisor feedback to employees | • Conflict resolution | • Chatbots for HR<br>• Employee file management<br>• Training for interactive learning |
| **Technology development** | • Brainstorming and ideation<br>• Development of new materials/ recipes<br>• Digital partner for ideas<br>• Generative algorithms for coding robots<br>• Identifying new topics | • Automating communication between departments for API deviations<br>• Design decisions<br>• Fraud detection<br>• Read and process technical specification documents |  | • Coding<br>• Editing technical designs<br>• Facilitate database search<br>• Scrape OBL results<br>• Software development<br>• Writing text, creating Publications and Patents |
| **Procurement** |  | • Review supplier contracts |  | • Purchasing processes |
| **Operations** | • Inventory management | • Identify bottlenecks<br>• Improvement of existing processes<br>• Process optimization through predictive quality |  | • General quality control<br>• Parameter tweaking<br>• Quality control and optimization through Image Processing (X-ray checks and visual inspections)<br>• Stabilization of (machine) workflows<br>• Standardization of payment processes |
| **Logistics** |  |  |  |  |
| **Marketing and sales** |  | • Pricing tactics<br>• Product portfolio more customer-orientated<br>• Recommendations for performance specifications from customers |  | • Customer relationship building<br>• Sales Assistance<br>• Standardization of the receipt process<br>• Translations<br>• Contact clients |
| **Service** |  |  |  | • Product performance monitoring |

handled with human oversight, as humans have the primary responsibility for employee development. Minimizing AI involvement in these areas is crucial due to the complex human and context-specific factors that AI cannot currently handle. Even when AI is involved, human judgments enriched with emotional intelligence are essential in arriving at the final decision.

Providing supervisor feedback to employees is another similar use case where AI seems inappropriate. Participants preferred receiving feedback directly from a human supervisor rather than AI. Human feedback is

The use of AI in decision-making and innovation is identified with mixed feelings due to concerns about the transparency of the process. Interview participants cited examples where AI has been successfully integrated into decision-making processes in some countries but emphasized the necessity for transparency. It is crucial to clearly identify and distinguish whether a human or AI makes a decision. This distinction helps to identify the sources used by AI and assess the level of bias involved in AI-generated outcomes.

Finally, client contact is also flagged as potentially risky for AI involvement due to customer preference for human interaction, which aligns with the company's cultural priorities. Maintaining personal interaction in customer communications is essential, especially since most of their customers are small businesses or individuals. This personal touch is valued over AI-managed interactions, highlighting the involvement of human touch in client contact.

## 5. Discussion

### 5.1. Interpretation of Results

Interview participants identified numerous use cases related to firm infrastructure and technology development. Organizations can consider these use cases in their prospective sensemaking to anticipate potential benefits in enhancing these value chain activities. This involves enhancing internal efficiencies through firm infrastructure and driving innovation through technology development to better support HAIC. Additionally, many identified use cases are related to the operations and Human Resource Management (HRM) activities of the value chain. This indicates that organizations can significantly realize benefits and value by adopting HAIC (Enholm et al., 2022) in these activities by enhancing productivity, improving decision-making, and enabling possible automation (Dwivedi et al., 2021). The AI use cases related to marketing and sales suggest that organizations are widely adopting in these areas due to their current and potential impact on business (Verma et al., 2021). We also observed AI use cases related to procurement and service and no use cases related to logistics. This may be because AI applications in these areas are still emerging, and organizations are still exploring AI use cases related to these activities. However, organizations can greatly benefit from AI applications in these areas of the supply chain (Toorajipour et al., 2021).

From a group task circumplex perspective, we observe many "choose" and "execute" tasks in the AI use cases. This richness of cases is expected, as AI excels at automating tasks, improving efficiencies, providing recommendations, and conducting data analysis (O' Leary, 2013; Trocin et al., 2021). organizations can prioritize these group tasks in their strategy when adopting HAIC.

Another observation is that HAIC use cases related to "generate" are few and still evolving (Verganti et al., 2020). We also observed one use case related to "negotiate" group tasks. Negotiation remains a complex task even for humans, involving considering emotions, context-specific judgments, relationship-building, and managing potential conflicts. Currently, AI is not capable of handling these complex human interactions, which explains the limited use cases in this area (Xu et al., 2020).

When looking at a holistic perspective of the value chain and group task circumplex combined, we see that many use cases are concentrated in the supporting activities of the value chain (technology development, firm infrastructure, and HRM). Most of the HAIC use cases within these supporting activities are related to "choose" group tasks. Therefore, the supporting activities can benefit significantly from the adoption of HAIC related to "choose" group tasks. Improved decision-making in these supporting activities can indirectly enhance the smooth functioning of the primary activities of the value chain as they receive better support (Piboonrungroj et al., 2017).

Regarding the primary activities of the value chain, many AI use cases are related to operations and marketing and sales, with a variety of cases falling into the "execute" group task category. Hence, the primary activities of the value chain may benefit more from adopting HAIC related to "execute" group tasks. Organizations looking to adopt HAIC in the future may consider leveraging HAIC in "choose" and "execute" group tasks within key value chain activities.

From a prospective sensemaking perspective, the identified use cases for HAIC not only serve to highlight current potential applications, but can also be used strategically to anticipate future developments (Jarrahi, 2018). Use cases provide insights into areas where AI could be increasingly used in the future and illustrate the potential impact on working methods and processes (Dwivedi et al., 2021). This enables companies to make informed decisions about future AI investments and strategies. In addition, exploring potential use cases encourages creative thinking and inspires the development of new ideas and innovations in the field of human-AI collaboration. By recognizing and anticipating future developments of HAIC at an early stage, companies can proactively prepare for change and actively shape it rather than simply reacting to it (Smith & Ashby, 2020).

The results of this study highlight the importance of anticipating potential AI impact on various organizational activities. The assignment of HAIC use cases along the value chain and the group task circle created a general understanding of the meaningful role of AI from the perspective of executives. This mapping supports the prospective sensemaking process by providing a clear, structured understanding of where AI can be most beneficial. It helps management to prioritize applications of AI in areas that deliver the highest benefits in terms of improved HAIC.

Our case study shows that the majority of HAIC use cases are centered around business infrastructure and technology development. For the sensemaking process, this means that the organization needs to identify these areas as strategic priorities. The limited AI use cases in Negotiate tasks highlight the current limitations of AI in handling complex human interactions. This insight is critical to the sensemaking process as it highlights the importance of balancing human and AI roles within the organization.

In addition, the results were also shared with the senior executive responsible for the case study company after the analysis and categorization of the use cases were finalized. The feedback is positive about the categorization of the cases. However, a further consideration of internal and external related value chain activities would be beneficial. Use cases that can be addressed internally can be managed and improved independently within the company. However, since HAIC solutions typically require integrating AI and production planning systems across different organization units, prospective sensemaking can differ for the same use case. External related use cases require collaboration with suppliers, customers, and other external partners. Here, divergent goals between the company and these partners may lead to hindering shared prospective sensemaking and, thus, difficulties in realizing use cases.

**5.2. Implications**

This study offers several implications for both theory and practice. First, we identified HAIC use cases and utilized these use cases as a potential tool for prospective sensemaking. The selected case organization is currently exploring how AI can be implemented to better support HAIC and make its processes more efficient. When conducting the semi-structured interviews and analyzing the results, we employed this prospective sensemaking approach to anticipate the potential of HAIC through the use cases. This approach involves individuals recognizing the feasibility of particular AI uses to support their work. If they perceive it as feasible, they are more interested in having it implemented as an AI solution to enhance their work through better HAIC. In this manner, we identified potential AI use cases to support prospective sensemaking. We identified 63 potential HAIC use cases, of which seven were flagged as potentially high risky for AI implementation.

Second, after identifying the use cases, we developed a novel matrix to categorize the identified 63 use cases. This contributes to theory by combining value chain activities and group task circumplex activities in a matrix to categorize the identified use cases. This approach helped us identify HAIC use cases related to various activities across the value chain and HAIC use cases covering different group task activities. It identifies the HAIC use cases separately for value chain activities and group task circumplex activities and demonstrates how they can support both activities in combination. For example, in the matrix, "Employee file management" is identified as a use case categorized under HRM as a value chain activity and execution as a group task circumplex activity. In combination, this is an HRM-execute activity, which helps us understand that this particular use case will support the better operation of the HRM value chain activity and enhance the execution of group tasks in HAIC. Hence, this combined approach provides a holistic view that encompasses value chain and group task activities, offering valuable contributions to theory and practice.

Third, with this holistic view of categorizing use cases by combining the value chain and group task circumplex, we observe that organizations recognize the potential of HAIC across different activities of the value chain (except for logistics) and different activities of the group task circumplex (Helo & Hao, 2022). This suggests a more balanced approach to exploring the prospective sensemaking of AI applications in supporting HAIC.

Fourth, the matrix provides the foundation for fostering collective sensemaking in the organization from a strategic perspective. Management can use the matrix as a map to help them decide where to start and invest resources most efficiently, as relevant members in the organization have prospective sensemaking. Hence, the mapping summarizes prospective sensemaking in the organization and provides an avenue for management to make sense of prioritizing future investments. This ensures that investments have the highest chance of fostering HAIC transformation.

The matrix of HAIC use cases, categorized by value chain activities and group tasks, serves as a valuable tool for practitioners for prospective sensemaking and strategic AI implementation. It provides a visual overview of potential AI applications, highlighting areas where AI can add the most value from a prospective sensemaking perspective and support decision making on where to invest resources. The categorization enables companies to proactively identify use cases for different areas of the business and ensure that their AI investments contribute to both operational efficiency and long-term strategic success.

Furthermore, while HAIC offers promising opportunities for organizations, the identified use cases, especially those involving email management and administrative tasks, highlight the critical importance of considering ethical implications and privacy concerns. Companies need to proactively address potential risks such as unintended biases in AI algorithms and the protection of sensitive personal data. Finding the right balance between the use of AI capabilities and compliance with ethical standards is essential for a responsible and sustainable adoption of AI.

Finally, we recognize the need for a broader discussion on the impact of AI on existing theoretical frameworks. Established frameworks such as Porter's Value Chain and McGrath's Group Task Circumplex are invaluable for understanding traditional organizational processes and group dynamics. However, they may require adaptation or extension to fully capture the unique complexities and opportunities introduced by human-AI collaboration. In this study, we address this g by combining both frameworks, thereby enhancing their applicability to the evolving landscape of human-AI interactions. AI continues to reshape the nature of work and value creation, future research could explore how these and other frameworks can be further developed to provide a more nuanced understanding of HAIC and its implications for management theory and practice.

## 6. Conclusion, Limitations and Outlook

This study investigates the implementation of HAIC within organizational settings and provides a comprehensive framework for mapping HAIC use cases across organizational value chain activities and group tasks to support prospective sensemaking.

This study has several limitations. Firstly, it is based on a limited number of interviews (14) from a global automotive company. This limits the generalizability of the results, as specific use cases and findings from this industry may not be transferable to other sectors with different challenges and requirements for AI applications. Secondly, the subjectivity of the semi-structured interviews influences the identification and categorization of HAIC use cases, which may lead to bias. Thirdly, AI technology rapidly evolves, so the identified use cases could quickly become outdated. Finally, potentially significant application areas such as logistics were not covered, suggesting that there may be untapped potential for AI applications in these areas. Further studies with a broader sample and in different industries are needed to get a more comprehensive picture of HAIC use cases and their impact.

Future research should consider several aspects to extend and deepen this study's findings. Firstly, this study's findings are based on a single automotive case company with a sample of 14 executives, which limits the generalizability to other industries and larger populations. While the insights gained are valuable within the current context, they may primarily serve as a guide for other industries. Future research could benefit from a broader and more diverse sample of organizations from different sectors to increase the generalizability of the results. Such a diverse sample could help to better understand the varying challenges and requirements of AI applications across different sectors, thereby enhancing the generalizability of the findings. Secondly, longitudinal studies could provide valuable insights into the evolving AI use cases and their long-term impact on organizational processes and performance. Such studies would make it possible to observe changes and developments in the use of AI over time. Finally, future studies should explore the potential role of AI in more complex and nuanced tasks such as negotiation and conflict resolution.

Currently, AI is still limited in these areas, but new possibilities could open up as technology advances. The development of AI systems capable of handling such challenging tasks could contribute significantly to the advancement of human-AI collaboration.

## Data Repository

Additional data of our study is stored here: https://doi.org/10.17605/OSF.IO/42RG7